\documentclass[epj]{svjour}
\usepackage{graphicx}% Include figure files
\usepackage{dcolumn}% Align table columns on decimal point
\usepackage{bm}% bold math
\usepackage{times}
\usepackage{latexsym}

\begin{document}
%\hfill TUM/T39-02-

\title{Nuclear many-body dynamics constrained by
QCD and chiral symmetry}
 
\author{P. Finelli\inst{1} \and
N. Kaiser\inst{2} \and
D. Vretenar\inst{3} \and
W. Weise\inst{2,4}
}

\institute{Physics Department, University of Bologna, and INFN - Bologna,
I-40126 Bologna, Italy 
%\email{paolo.finelli@bo.infn.it}
\and 
Physik--Department, Technische Universit\"at M\"unchen,
D-85747 Garching, Germany
%\email{Norbert.Kaiser@Physik.TU-Muenchen.DE}
\and 
Physics Department, Faculty of Science, University of
Zagreb, Croatia
%\email{vretenar@phy.hr}
\and 
ECT$^*$, I-38050 Villazzano (Trento), Italy
%\email{weise@ect.it}
}

\abstract{
We present a novel description of nuclear many-body systems, both for nuclear
matter and finite nuclei, emphasizing the connection with the condensate
structure of the QCD ground state and spontaneous chiral symmetry
breaking. Lorentz scalar and vector mean-fields are introduced in accordance
with QCD sum rules. Nuclear binding arises from pionic fluctuations, 
using in-medium chiral perturbation theory up to three-loop order. Ground state
properties of $^{16}$O and $^{40}$Ca are calculated. The built-in QCD
constraints reduce the number of input parameters significantly in comparison
with purely phenomenological relativistic mean-field approaches.
} 
\bigskip

\PACS{12.38.Bx, 21.65.+f, 21.60.-n, 21.30.Fe}

\maketitle
 
\section{Introduction}

The description of nuclear many-body dynamics must ultimately be constrained by
the underlying theory of the strong interaction -- Quantum Chromodynamics (QCD). 
Previous phenomenological steps with this goal in mind have been taken by 
Quantum Hadrodynamics (QHD)~\cite{SW.97}. In the mean field (Hartree)
approximation, such an approach is equivalent to a model with local four-point
interactions between nucleons~\cite{MAD.92,RUS.98,Bur.01}. 
Models based on QHD have been successfully 
applied to describe a variety of nuclear phenomena over the whole periodic
table, from light nuclei to superheavy elements (see Ref.~\cite{Rin.96} for a
recent review, and references therein).

While this phenomenological success is impressive, an understanding of its
foundations in QCD is still missing. The multitude of input parameters in QHD
models is usually not constrained by QCD considerations. Explicit pionic
degrees of freedom are absent in most QHD type calculations, whereas it is
obvious that pions, as Goldstone bosons of spontaneously broken chiral
symmetry, must play an important role in the nuclear many-body
problem. Two-pion exchange effects are supposedly incorporated as part of
the strong scalar-isoscalar field of QHD models, but in an ad-hoc manner
without detailed reference to the underlying $\pi\pi NN$ dynamics.
 
A general low-energy effective Lagrangian for nuclear
systems can been written down as a Taylor series in point couplings involving
nucleon currents and their derivatives~\cite{FML.96,FS.00}. 
A large number of coefficients
must be determined in such an effective field theory. The empirical data
set of nuclear bulk and single-particle properties can be used to fix no
more than six or seven of these parameters. Our approach is similar in 
spirit but proceeds with a different strategy, imposing as many QCD
constraints as possible in order to minimize the number of free parameters. 

The success of relativistic mean-field phenomenology has been attributed
primarily to large Lorentz scalar and vector 
nucleon self-energies~\cite{FS.00}. There is
evidence, in particular from nuclear matter saturation and from spin-orbit
splittings in finite nuclei, that the magnitudes of these scalar and vector
potentials are of the order of several hundred MeV in the nuclear
interior. Investigations based on QCD sum rules~\cite{DL.90,CFG.91,Jin.94} have
shown how such large scalar and vector nucleon self-energies arise in
finite-density QCD, at least qualitatively, through changes in the quark
condensate and the quark density. Such QCD sum rule constraints will be one of
the basic elements of our discussion.

The second important ingredient is chiral pion-nucleon dynamics. In
Ref.~\cite{KFW.01} the equation of state of isospin-symmetric 
nuclear matter has been calculated using in-medium chiral perturbation theory. 
At nuclear matter saturation density, the 
Fermi momentum $k_f$ and the pion mass $m_\pi$ represent comparable scales, and
therefore pions must be included as explicit degrees of freedom in the 
description of nuclear many-body dynamics. The calculations have been performed
to three-loop order and incorporate the one-pion exchange Fock term, iterated
one-pion exchange and irreducible two-pion exchange. The resulting nuclear 
matter equation of state is expressed as an expansion in powers of the Fermi 
momentum $k_f$. The expansion coefficients are functions of $k_f/m_\pi$, the 
dimensionless ratio of the two relevant scales. The calculation involves one 
single parameter, the momentum space cutoff $\Lambda$ which encodes 
NN-dynamics at short 
distances. With  $\Lambda \simeq  0.65$ GeV adjusted 
to the energy per particle $\bar E(k_{f0}) = -15.3$ MeV,
the calculated equation of state gives the density  
$\rho_0=0.178$\,fm$^{-3}$, the compression modulus $K = 255$ MeV, 
and the asymmetry energy 
$A(k_{f0}) = 33.8$\,MeV at saturation.

Based on these observations, our ``minimal'' approach for nuclear matter and 
finite nuclei starts from the following hypothesis: 

{\bf A)} The nuclear matter ground state is characterized by large 
scalar and vector nucleon self-energies of approximately equal magnitude
and opposite sign, arising from the in-medium change 
of the scalar quark condensate and the quark density.

{\bf B)} Nuclear binding and saturation result from chiral (pionic)
fluctuations superimposed on the condensate background fields. These
pionic fluctuations are calculated according to the rules of in-medium
chiral perturbation theory.

As concerns hypothesis A), finite-density QCD sum rules 
\cite{DL.90,CFG.91,Jin.94} predict the scalar and vector potentials
to be each about $300 - 400$ MeV 
in magnitude at nuclear matter saturation density $\rho_0$. The same
QCD sum rule analysis, taken to leading order, also suggests the ratio 
of scalar to vector fields to be close to $-1$. We shall argue that this
is indeed a valid starting point, though not yet capable of producing 
nuclear binding. Hypothesis B) asserts that binding and saturation is ruled
primarily by explicit $\pi\pi$ exchange dynamics based on known properties
of $\pi N$ interactions and calculable using systematic methods of chiral
effective field theory - at least as long as the Fermi momentum $k_F$
is small compared to the characteristic scale, $4\pi f_{\pi}\sim 1~GeV$,
associated with spontaneous chiral symmetry breaking in QCD.

Our aim in this paper is thus to study the interplay between condensate
background fields and perturbative chiral fluctuations, both rooted in the
spontaneous symmetry breaking pattern of QCD, in forming nuclei. We will
demonstrate that this scenario works at large, once a single scale parameter
is set to reproduce nuclear matter at equilibrium. Whereas in first
approximation the condensate potentials do not play a role for the 
saturation mechanism, we will show that they
are essential for the description of ground states of finite nuclei. We
restrict ourselves  here to gross features of isospin-symmetric $(N=Z)$ 
nuclei and relegate further fine tunings as well as the $N>Z$
case to forthcoming work. 

\section{Model for nuclear matter and finite nuclei}

\subsection{Lagrangian}

Our approach is defined by the following
(isoscalar) Lagrangian, relevant for $N=Z$ nuclei:
\begin{eqnarray}
\cal{L} & = & \bar{\psi}(i\gamma_{\mu}\partial^{\mu}-M)\psi \nonumber\\
 & & +\frac{1}{2}G_{S}(\rho) \,\bar{\psi}\psi \,\bar{\psi}\psi
-\frac{1}{2}G_{V}(\rho)\,\bar{\psi}\gamma_{\mu}\psi\, \bar{\psi}\gamma^{\mu}
\psi \nonumber \\& & +\frac{1}{2}D_{S}(\rho)\, \partial_{\nu}\bar{\psi}\psi\,
\partial^{\nu}\bar{\psi}\psi - \frac{1}{2}D_{V}(\rho)\, \partial_{\nu}\bar{
\psi}\gamma_{\mu}\psi\,\partial^{\nu}\bar{\psi}\gamma^{\mu}\psi \nonumber \\
& & +{e\over 2}A^{\mu}\bar{\psi}(1+\tau_3)\gamma_{\mu}\psi - \frac{1}{4}
F_{\mu\nu}F^{\mu\nu} \,.
\label{lagrangian}
\end{eqnarray}
Here, $\psi$ is the nucleon spinor field, $M$ is the (free) nucleon mass and 
the subscripts $S$ and $V$ refer to the scalar and vector type interactions, 
respectively. The vector potential and field strength tensor of the 
electromagnetic field are denoted $A^{\mu}$ and $F^{\mu\nu}$. The coupling
parameters of the four-nucleon contact interactions and the derivative terms 
are assumed to be functions of the nucleon density $\rho$. These coupling
strengths include contributions from condensate background fields and
pionic (chiral) fluctuations, to be specified. We will formally work at the 
mean field level using Eq.(1), with the understanding that fluctuations
beyond mean field are encoded in the density dependent coupling strengths.

The single-nucleon Dirac equation derived 
from the Lagrangian Eq. (\ref{lagrangian}) by
variation with respect to $\bar \psi$, reads:
\begin{equation}
[\gamma_{\mu}(i\partial^{\mu} - \Sigma^{\mu} 
-\Sigma_{R}^{\mu}) - (M + \Sigma_s + \Sigma_{Rs})]\psi = 0 \,,
\label{Dirac}
\end{equation}
with the nucleon self-energies defined by the following relations:
\begin{eqnarray}
\Sigma^{\mu} & = & G_V j^{\mu} - D_V \Box j^{\mu} - 
eA^{\mu}\frac{1+\tau_3}{2}\\
\Sigma_s & = & - G_S (\bar{\psi} \psi) + D_S \Box (\bar{\psi} \psi)\\
\Sigma_{Rs} &  = & \frac{\partial D_S}{\partial \rho} (\partial_{\nu}
j^{\mu}) u_{\mu} (\partial^{\nu} (\bar{\psi} \psi))\\
\Sigma_R^{\mu} & = & \left( - \frac{1}{2}  
\frac{\partial G_S}{\partial \rho} (\bar{\psi} \psi)
(\bar{\psi} \psi) - \frac{1}{2}  
\frac{\partial D_S}{\partial \rho} (\partial^{\nu} (\bar{\psi} \psi))
(\partial_{\nu} (\bar{\psi} \psi)) \right. \nonumber\\
& ~ & + \left. \frac{1}{2} \frac{\partial G_V}{\partial \rho} j^{\nu}
j_{\nu} + \frac{1}{2} \frac{\partial D_V}
{\partial \rho}(\partial_{\nu} 
j_{\alpha})(\partial^{\nu} j^{\alpha}) \right) u^{\mu}\nonumber\\
& ~ & - \frac{\partial D_V}{\partial \rho} 
 (\partial_{\nu} j_{\alpha}) u^{\alpha} (\partial^{\nu}j^{\mu}) \,, 
\end{eqnarray}
where $j^{\mu} = \bar{\psi}\gamma^{\mu}\psi$ is the nucleon current, 
and the velocity $u^{\mu}$ is defined by $\rho u^{\mu} = j^{\mu}$. In 
addition to the usual vector $\Sigma^{\mu}$ and scalar $\Sigma_s$ 
self-energies, the density dependence of the vertex functions
$G_S(\rho)$, $G_V(\rho)$, $D_S(\rho)$ and $D_V(\rho)$,  
produces the {\it rearrangement} contributions $\Sigma_{Rs}$ and 
$\Sigma_R^{\mu}$~\cite{FLW.95}. The inclusion of the rearrangement
self-energies is essential for energy-momentum conservation and the 
thermodynamical consistency of the model~\cite{FLW.95,TW.99}.

The ground state of a nucleus with A nucleons
is the product of the lowest occupied single-nucleon
self-consistent stationary solutions of the Dirac equation Eq.(\ref{Dirac}).
The ground state energy is the sum of the single-nucleon energies plus
a functional of the scalar density
\[
\rho_S = \sum\limits_{k=1}^{A} \bar \psi_k \psi_k \,
\]
and of the nucleon (vector) density 
\[
\rho = \sum\limits_{k=1}^{A} \psi^{\dagger}_k \psi_k \,
\]
calculated in the {\it no-sea} approximation, i.e. the 
sum runs only over occupied positive-energy single-nucleon states with 
wave functions $\psi_k$.

\subsection{Nuclear matter}

The energy density $\cal E$ and the pressure $P$ of isospin symmetric nuclear
matter are calculated from the energy-momentum tensor $T^{\mu \nu}$ as 
\begin{equation}
\label{ENM}
{\cal E}  =  {\cal E}_0    + \frac{1}{2}G_S \rho_S^2
   + \frac{1}{2}G_V \rho^2 \,,
\end{equation}

\begin{eqnarray}
\label{Press}
P = \rho {\partial {\cal E} \over \partial \rho} - 
{\cal E}  & = & \mu^* \rho - {\cal E}_0 + \frac{1}{2}G_V \rho^2 - 
   \frac{1}{2} G_S \rho_S^2 \nonumber\\    
  & ~ &  - \frac{1}{2}    \frac{\partial G_S} {\partial \rho} \rho_S^2 \rho +
   \frac{1}{2} \frac{\partial G_V}{\partial \rho} \rho^3 \,. 
\end{eqnarray}

It should be pointed out that, while these expressions are formally
derived in the mean field (Hartree) approximation from the Lagrangian (1),
they incorporate exchange effects and fluctuations beyond mean field.
In particular, the pionic fluctuations to be described in more detail
in section 2.4 are calculated at three-loop order which includes Fock
terms from one-pion exchange as well as all possible exchange terms
related to two-pion exchange. These effects are transcribed into the 
density dependence of the couplings $G_{S,V}(\rho)$. 

The free quasi-particle contribution is given by 
\begin{equation}
{\cal E}_0 = \frac{4}{(2\pi)^3} \int\limits_{|\vec k\,|\le k_f}
 d^3 k \sqrt{ k^2 + M^{*2}} =  \frac{1}{4} ( 3 \mu^* \rho +M^* \rho_S)\,,
\end{equation}
with the effective chemical potential 
\begin{equation}
\mu^* = \sqrt{k_f^2 + M^{*2} }\,,
\end{equation}
and the effective nucleon mass 
\begin{equation}
M^* = M - G_S \rho_S \,.
\end{equation}

The baryon density is related to the Fermi momentum $k_f$ in the usual way, 
$\rho = 2k_f^3/3\pi^2$, and the expression for the scalar density reads
\begin{eqnarray}
\label{SCADENS}
\rho_S & = & \frac{4}{(2\pi)^3} \int\limits_{|\vec k|\le k_f}
d^3 k \frac{M^*}{\sqrt{ k^2 + M^{*2}}} = \nonumber\\
 & ~ & \frac{M^*}{\pi^2} \left[ k_f \mu^*
- M^{*2} \ln \frac{k_f + \mu^*}{M^*}    \right] \,. 
\end{eqnarray}
Note that, in contrast to the energy density, {\it rearrangement} contributions
appear explicitly in the expression for the pressure.

The general form of the vertex functions $G_S(\rho)$ and $G_V(\rho)$ is
\begin{eqnarray}
G_S(\rho) & = & G_S^{(0)} - \Delta G_S(\rho)\\
G_V(\rho) & = & G_V^{(0)} + \Delta G_V(\rho) \quad ,
\end{eqnarray}
where $G_{S,V}^{(0)}$ are terms governed by the QCD condensates, and $\Delta 
G_{S,V}(\rho)$ refer to the pionic fluctuations, re-expressed as density 
dependent corrections to the mean-fields. 

\subsection{Constraints from QCD condensates}

The in-medium QCD sum rules relate the changes in the scalar 
quark condensate and the quark density due to the finite baryon
density with the scalar and vector self energies of a nucleon in 
the nuclear medium. In leading order, which should be valid below
and around nuclear matter saturation density, one finds for these
condensate parts of the nucleon self energies~\cite{CFG.91}:
\begin{equation}
\Sigma_S^{(0)} = - \frac{8 \pi^2}{\Lambda_B^2} 
(<\bar q q >_{\rho} - <\bar q q >_{\rm vac}) = 
- \frac{8 \pi^2}{\Lambda_B^2} \frac{\sigma_N}{m_u + m_d}~\rho_S
\end{equation}

\begin{equation}
\Sigma_V^{(0)} = \frac{64 \pi^2}{3 \Lambda_B^2} <q^{\dagger} q >_{\rho}~
        = \frac{32 \pi^2}{\Lambda_B^2}~\rho \quad ,
\end{equation}
where $\Lambda_B \approx 1$ GeV is a characteristic scale, the Borel mass,
entering in the QCD sum rule analysis. For typical
values of the nucleon sigma term $\sigma_N$ and the current quark masses
$m_u$ and $m_d$, the ratio 
\begin{equation}
\frac{\Sigma_S^{(0)}}{\Sigma_V^{(0)}} = - 
\frac{\sigma_N}{4 (m_u + m_d)}
\end{equation}
is close to $-1$ (take, for example, $\sigma_N \simeq 45\,$MeV and $m_u+m_d
\simeq 12\,$MeV), with uncertainties at the $20\%$ level.
Using the Hellmann-Feynman theorem in combination with PCAC 
(the Gell-Mann-Oakes-Renner relation) to derive the in-medium scalar 
quark condensate, one finds in the Fermi gas approximation~\cite{DL.90}:

\begin{equation}
\Sigma_S^{(0)} = M^* - M = - \frac{\sigma_N M}
{m_\pi^2 f_\pi^2}~\rho_S \quad ,
\end{equation}
which implies:
\begin{equation}
\Sigma_S^{(0)} (\rho_0)\simeq -350~{\rm MeV}~\frac{\sigma_N}
{50~{\rm MeV}} \,{\rho_S\over \rho_0} 
\end{equation}
or, identifying $\Sigma_S^{(0)} = - G_S^{(0)}\rho_S$:
\begin{equation}
\label{estimate} 
G_S^{(0)} \simeq  11~{\rm fm}^2~\frac{\sigma_N}{50~{\rm MeV}} \quad at \,\,\,\rho_S
\simeq \rho_0 = 0.16~{\rm fm}^{-3}.
\end{equation}

Up to this point our discussion is based on the leading terms of the
QCD sum rule, the ones involving the dimension-four condensates 
$\langle m_q \bar{q}q \rangle$ and $\langle G_{\mu\nu}G^{\mu\nu} \rangle$.
The density dependence of the gluon condensate is in fact weak and need
not be considered. The influence of higher-dimensional condensates has
been discussed in great detail in Ref.~\cite{CFG.91}. Uncertainties
arise primarily from contributions of four-quark condensates. It is common
practice to approximate those four-quark condensates assuming factorization
which introduces potentially large and uncontrolled errors. We can refrain
from this discussion because our {\it explicit} treatment of scalar $\pi\pi$
fluctuations removes at least part of these uncertainties.

\subsection{Pionic (chiral) fluctuations}

This brings us next to the constraints from chiral pion-nucleon dynamics
on the density dependent parts of Eqs.(13,14).
If we follow the assumption, also made implicitly in Ref.~\cite{KFW.01},
that in nuclear matter $\Sigma_S^{(0)} \simeq - \Sigma_V^{(0)}$ 
at $\rho= \rho_0$, the density 
dependent couplings of the pionic fluctuation terms
$\Delta G_S(\rho)$ and $\Delta G_V(\rho)$ are determined by equating
the corresponding self-energies in the single-nucleon Dirac equation
(\ref{Dirac}) with those calculated using 
in-medium chiral perturbation theory (CHPT) in Ref.[11]:
\begin{equation}
\label{PCC1}
\Delta G_S(\rho) \rho_S = \Sigma_S^{\rm CHPT} (k_f,\rho)~~~~,
\end{equation}
\begin{equation}
\label{PCC2}
\Delta G_V(\rho) \rho + \frac{1}{2}  
\frac{\partial \Delta G_S}{\partial \rho} \rho_S^2 +
\frac{1}{2}  
\frac{\partial \Delta G_V}{\partial \rho} \rho^2 =
\Sigma_V^{\rm CHPT} (k_f,\rho) ~~~.
\end{equation}
We have indicated here that the $\Sigma^{CHPT}_{S,V}(p,\rho)$ depend explicitly
on the nucleon momentum $p$.

\begin{figure}
\rotatebox{-90}{\resizebox{0.40\textwidth}{!}{%
  \includegraphics{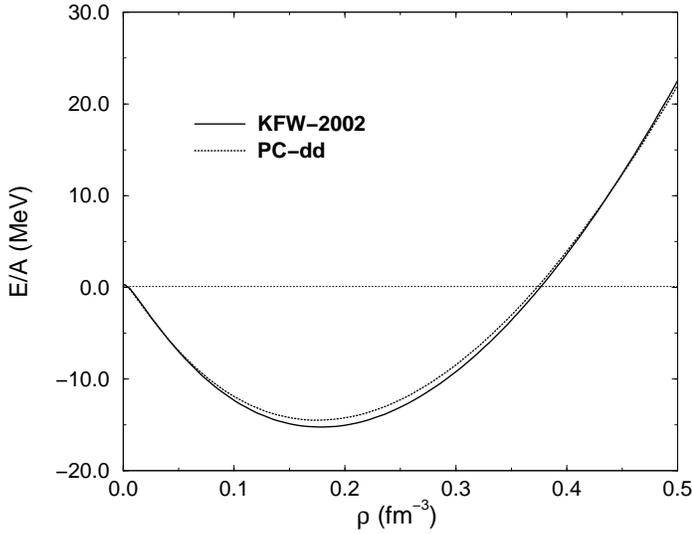}}}
\caption{\label{fig1}Binding energy per nucleon for symmetric nuclear
matter as a function of the baryon density. The solid curve (KFW-2002)
is the EOS calculated in Ref.~\protect\cite{KFW.01} by using
in-medium CHPT. 
The EOS displayed by the dotted curve (PC-dd) is obtained when the resulting 
CHPT nucleon potentials are mapped on the self-energies
of the relativistic point-coupling model with density dependent
couplings.}
\end{figure}

The energy per particle 
$\bar E(k_f)$ in nuclear matter gives,
via the Hugenholtz-van Hove theorem, the sum of the scalar and vector nucleon 
self-energies $U(k_f,k_f) = \Sigma_S^{\rm CHPT}(k_f,\rho)+ \Sigma_V^{\rm CHPT}
(k_f,\rho)$  at the Fermi surface $p=k_f$ up to two-loop order, 
as generated by 
chiral one- and two-pion exchange \cite{KFW.02}. 
The difference $\Sigma_S^{\rm CHPT}(p,\rho)- \Sigma_V^{\rm CHPT} (p,\rho)$ is
calculated from the same pion-exchange diagrams via the anti-nucleon single
particle potential in nuclear matter. Following a procedure similar to the 
determination of the nucleon-meson vertices of relativistic mean-field models 
from Dirac-Brueckner calculations \cite{HKL.01}, we neglect the momentum 
dependence of $\Sigma_{S,V}^{\rm CHPT}(p,\rho)$ and take their values at the 
Fermi surface $p=k_f$. A polynomial fit up to order $k_f^5$ is performed, 
and the self-energies are then reexpressed in terms of baryon 
density $\rho = 2k_f^3/3\pi^2$: 
\begin{eqnarray}
\label{SSE}
\Sigma_S^{\rm CHPT} (k_f,\rho) & = &  (c_{S0} + c_{S1}~\rho^{1/3} +
c_{S2}~\rho^{2/3} )~\rho \, \\
\label{VSE}
\Sigma_V^{\rm CHPT} (k_f,\rho) & = & (c_{V0} + c_{V1}~\rho^{1/3} +
c_{V2}~\rho^{2/3} )~\rho \,. 
\end{eqnarray}
The values of the coefficients are: $c_{S0} = - 2.805 ~{\rm fm}^2$, 
$c_{S1} = 2.738 ~{\rm fm}^3$, $c_{S2} = 1.346 ~{\rm fm}^4$,
$c_{V0} = -2.718 ~{\rm fm}^2$, $c_{V1} = 2.841 ~{\rm fm}^3$, 
and $c_{V2} = 1.325~{\rm fm}^4 $. The resulting expressions for the 
density dependent couplings of the pionic fluctuation terms are
\begin{equation}
\Delta G_S(\rho) = c_{S0} + c_{S1}~\rho^{1/3} +
c_{S2}~\rho^{2/3}
\label{GS}
\end{equation}
\begin{equation}
\Delta G_V(\rho) = c_{V0} + 
\frac{1}{7} (6 c_{V1} - c_{S1})~\rho^{1/3} +
\frac{1}{4} (3 c_{V2} - c_{S2})~\rho^{2/3} \,.
\label{GV}
\end{equation}
%%%%%%%%%%%%%%%%%%%%%%%%%%%%%%%%%%%%%%%%%%%%%%%%%%%%%%%%%%%%%%%%%%%%%%%%%%%
In deriving the expressions for $\Delta G_S(\rho)$ and $\Delta G_V(\rho)$ 
we have set $\rho_S \approx \rho$ on the left hand sides of
of Eqs.~(\ref{PCC1}) and (\ref{PCC2}). Although the relation between the scalar
and baryon density depends on the Fermi momentum, this approximation
is justified for $\rho \leq \rho_0$.
With the density dependent couplings (\ref{GS}) and (\ref{GV}), 
Eqs. (\ref{ENM}) to (\ref{SCADENS}) produce a nuclear matter equation 
of state which is very close to the one calculated successfully in CHPT.
This is shown in Fig.~\ref{fig1} where we compare the 
nuclear matter equation of state calculated from 
Eqs. (\ref{ENM}) and (\ref{Press}), with the one obtained 
using in-medium CHPT ~\cite{KFW.01}. Corresponding
ground state properties:
the binding energy per particle, the saturation density,
the compressibility modulus,
and the asymmetry energy at saturation, are compared in 
Table~\ref{tab1}. Small differences arise mainly because the 
momentum dependence of
the CHPT self-energies has been frozen in Eqs.(23,24). This is a well known 
problem which has been extensively discussed, for instance, 
in Ref.~\cite{HKL.01}. By fine tuning just two of the parameters
in Eqs. (\ref{GS}) and (\ref{GV}) we could, of course, reproduce the
CHPT equation of state of Ref.~\cite{KFW.01} exactly. In the 
present work, however, we prefer not to perform any such tuning of
parameters. 

\subsection{Finite nuclei}

Having constrained $G^{(0)}_{S,V}$ from QCD condensates, and having
adjusted one short-distance scale parameter appearing in the pionic 
fluctuation couplings $\Delta G_S(\rho)$ and $\Delta
G_V(\rho)$ to the equation of state of isospin symmetric nuclear matter, we
proceed to calculate finite nuclei. In this work we only consider the isoscalar
channel and calculate the ground states of $^{16}$O and $^{40}$Ca.
In addition to $G_S(\rho)$ and $G_V(\rho)$, two new quantities
appear specifically for finite nuclei, namely the couplings of the terms
involving derivatives in the nucleon fields in Eq.(\ref{lagrangian}): 
$D_S(\rho)$ and $D_V(\rho)$. Guided by 
dimensional considerations, we introduce the ansatz

\begin{equation}
D_S(\rho) = \frac{G_S(\rho)}{\Lambda^2}\quad{\rm and} \quad 
D_V(\rho) = \frac{G_V(\rho)}{\Lambda^2}\quad ,
\end{equation}
where $\Lambda$ is again a characteristic mass scale delineating 
short and long distance phenomena. In the present calculation
we simply choose $\Lambda \approx 0.65$ GeV, the same value that has 
been used for the 
momentum space cutoff in the in-medium CHPT calculation of 
the nuclear matter equation
of state. In this way, and we emphasize this point, no new 
parameters are needed in the calculation of finite nuclei. 

\begin{figure}
\rotatebox{-90}{\resizebox{0.40\textwidth}{!}{%
  \includegraphics{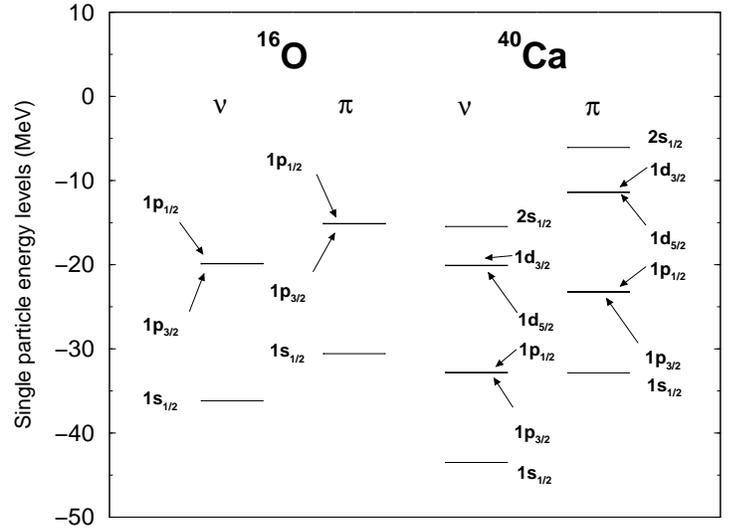}}}
\caption{\label{fig2} Neutron and proton single-particle levels
in $^{16}$O and $^{40}$Ca calculated in the relativistic point-coupling model.
The calculation is performed using only the contribution 
from chiral one- and two-pion exchange
to the density dependence of the coupling parameters (i.e. $G_{S,V}^{(0)}=0$ in
Eqs.(13,14)).}
\end{figure}

In the first step we have calculated the ground states of  
$^{16}$O and $^{40}$Ca with $G_S(\rho) = \Delta G_S(\rho)$
and $G_V(\rho) = \Delta G_V(\rho)$, i.e. we have set the couplings 
to the condensate background fields to zero. The nuclear dynamics
is then completely determined by chiral (pionic) fluctuations.
The interesting result is that the calculated total binding energies
are within $5-8 \%$ of the experimental values, but the resulting radii 
of the two nuclei are too small (by about 0.2 fm). This is because
the spin-orbit partners $(1p_{3/2},1p_{1/2})$ and
$(1d_{5/2},1d_{3/2})$ are practically degenerate: chiral two-pion exchange
dynamics alone, although it can provide the attraction necessary to bind
nuclei, does not produce the proper spin-orbit interaction.
This is shown in Fig.~\ref{fig2}, where we display the neutron
and proton single-particle levels in $^{16}$O and $^{40}$Ca, calculated
in the limit $G_{S,V}^{(0)} = 0$.

The spin-orbit degeneracy is removed by 
including the self-energies which arise from the changes in 
the scalar quark condensate and quark density. In the second step
we have adjusted the couplings $G_S^{(0)}$ and $G_V^{(0)}$ to
the binding energies and 
the charge radii of $^{16}$O and $^{40}$Ca. We emphasize again that, up
to this point, our calculation of both the nuclear matter
equation of state and the binding energies of finite nuclei includes
only one adjustable parameter: $\Lambda = 0.647$ GeV. Even though
$G_S^{(0)}$ and $G_V^{(0)}$ were varied independently, the 
minimization procedure tends to favour cancellation 
of the contributions from the
corresponding large scalar and vector self-energies. This happens 
because there is already enough binding from pionic fluctuations,
and therefore $\Sigma_S^{(0)} = - \Sigma_V^{(0)}$ represents a 
very good approximation for the condensate potentials. The final values
$G_S^{(0)} =  10.52~{\rm fm}^2$ and $G_V^{(0)} = 10.00~{\rm fm}^2$
should be compared with the estimate Eq.(\ref{estimate}), and with 
the leading order coefficients of the pionic (CHPT) terms 
$c_{S0}$ and $c_{V0}$. This is a remarkable result which indeed supports
the ``minimal scenario'' with condensate background fields plus pionic
fluctuations as a very reasonable starting point. 

\begin{figure}
\rotatebox{-90}{\resizebox{0.40\textwidth}{!}{%
  \includegraphics{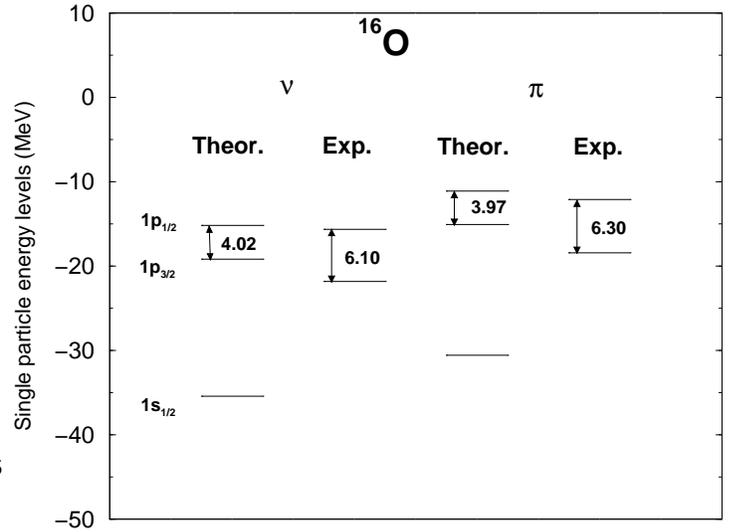}}}
\caption{\label{fig3} The neutron and proton single-particle levels
in $^{16}$O calculated in the relativistic point-coupling model, are shown
in comparison with experimental levels. The calculation is performed
by including both the contributions of chiral pion-nucleon exchange, and 
of the isoscalar condensate self-energies,
to the density dependence of the coupling parameters.}
\end{figure}

In Table \ref{tab2} we compare the calculated binding energies and charge 
radii with the corresponding empirical values. The absolute deviations
between theory and experiment are 7.8\%
and 5.3\% for the binding energies, and 2.5\% and 4.6\% for the charge radii of
$^{16}$O and $^{40}$Ca, respectively.  
In Fig.~\ref{fig3} the neutron
and proton single-particle levels in $^{16}$O, calculated 
with the inclusion of the condensate potentials, are   
compared with the experimental levels. We notice that this calculation 
reproduces about $2/3$ of the 
empirical spin-orbit splitting. While this result is clearly not yet 
satisfactory at a quantitative level, it also indicates the necessary
steps for further improvements. The approach is so far not fully
self-consistent, in the following sense. By construction, the condensate
potentials do not contribute to the binding of nuclear matter which is
accounted for almost entirely by chiral two-pion exchange dynamics. This
leaves no room for increasing the scalar $(S)$ and vector $(V)$ condensate
background contributions such that, by their difference $S - V$, 
the large spin-orbit
splitting can be reproduced quantitatively. The obvious solution is to
treat the chiral (two-pion exchange) fluctuations and the condensate
self-energies on the same level, both for nuclear matter and for 
finite nuclei. Furthermore, it is necessary 
to go beyond the leading order in the QCD sum rules and examine higher 
order density dependence for the condensate self-energies.
These points will be considered
in a forthcoming analysis of a generalized point-coupling model
constrained by QCD sum rules and in-medium CHPT.        

We note in passing that our model for finite nuclei could, in 
principle, be based on an alternative chiral approach to 
isospin-symmetric nuclear matter proposed by Lutz {\it et al}~\cite{Lutz.00}.
In addition to pion-exchange, 
their approach includes contributions from a zero-range
NN-contact interaction treated beyond the mean-field approximation
(i.e. the contact interaction is also iterated with $1\pi$ exchange). However,
while the nuclear matter equation of state of Ref.~\cite{Lutz.00} 
is comparable to the one used in the present work, the single-nucleon
potential resulting from that approach has several unrealistic 
features~\cite{FK.02}. For instance, the single-nucleon potential
at zero momentum $U(0,k_{f0}) \approx -23$ MeV is not sufficiently
attractive,
and the total single-nucleon energy does not increase monotonically 
with momentum (implying a negative effective mass).
The corresponding scalar and vector self-energies resulting from this
particular scheme would therefore not be
suitable for applications to finite nuclei. 

\section{Conclusion}

The effective Lagrangian Eq. (\ref{lagrangian}), 
with couplings governed by scales of low-energy QCD, gives a good
description of both symmetric nuclear matter and finite $N=Z$ nuclei
at a level better than 10\%, even without detailed fine tuning of 
parameters. While nuclear binding and 
saturation are almost completely generated by chiral (two-pion exchange)
fluctuations, strong scalar and vector fields of equal magnitude and
opposite sign, induced by changes of the QCD vacuum in the presence 
of baryonic matter, drive the large spin-orbit splitting in finite nuclei.
Considering the constraints 
from QCD condensates and chiral dynamics that keep the number of
adjustable parameters at minimum, our results are quite encouraging.
Investigations are now being generalized to include corrections from 
higher dimensional QCD condensates. The calculations are also expanded to 
cover a wider range of finite nuclei with extensions towards $N > Z$ systems.

\begin{acknowledgement}
This work has been supported in part by BMBF, GSI, DFG and INFN.
\end{acknowledgement}

\newpage

\begin{table}
\begin{center}
\caption{Nuclear matter saturation properties: binding energy 
per nucleon, saturation density, incompressibility, and
the asymmetry energy at saturation. The first row corresponds
to the in-medium CHPT calculation including one- and two-pion 
exchange ~\protect\cite{KFW.01}. The EOS displayed in the second row
is obtained when the resulting 
CHPT single nucleon potentials are mapped on the self-energies
of the relativistic point-coupling model with density dependent
couplings.}
\bigskip
\begin{tabular}{c c c c c}
\hline
\hline
   {\sc model} & E/A (MeV) & $\rho_0$ ({\rm $fm^{-3}$}) & K (MeV) 
   & $A$ (MeV)\\
\hline
CHPT \protect\cite{KFW.01}& -15.26 & 0.178 & 255 & 33.8 \\
PC-DD & -14.51 & 0.175 & 235 & 36.6 \\
\hline
\hline
\end{tabular}
\label{tab1}
\end{center}
\end{table}

\begin {table}
\caption{Binding energies per nucleon E/A [MeV] and root mean square
charge radii r$_c$ [fm] of $^{16}$O and $^{40}$Ca. 
The experimental values, shown in the second and third column,
are compared with the results of the present calculation.}
\label{tab2}
\begin {center}
\begin{tabular}{c c c c c c c c}
\hline
\hline
  &E$^{exp}$/A & & r$_c^{exp}$ & & E/A & & r$_c$\\
\hline
$^{16}$O &7.98 & &2.74 & &8.60 & &2.80 \\
$^{40}$Ca &8.55 & &3.48 & &8.10 & &3.64 \\
\hline
\hline
\end {tabular}
\end{center}
\end{table}

\end{document}